\documentclass{icrc}
\usepackage{times}
\usepackage{graphicx} 
\begin{document}
\title{A Measurement of Secondary Muon Angular Distributions}
\author[1]{J. Poirier}
\author[1]{C. D'Andrea}
\affil[1]{Center For Astrophysics at Notre Dame, Physics Dept., University of Notre Dame, 
Notre Dame, Indiana 46556 USA}
\correspondence{John Poirier (Poirier@nd.edu)}
\runninghead{Poirier: Asymmetries}
\firstpage{1}
\pubyear{2001}
\maketitle
\begin{abstract}
Project GRAND is an array of proportional wire chambers composed
of 64 stations with each station containing eight proportional
wire planes and a 50 mm steel plate.  The proportional wire planes
together with the steel absorber allow a measurement of the angle and
identity of single muon tracks.  Of the two data-taking
triggers, one is for single tracks.  Since the rate of single muon
tracks at sea level above the muon energy threshold of the
experiment (0.1 GeV) is substantial, good statistics are available by
accumulating several years of data.  A map in sidereal right ascension and
declination is obtained as well as one in a sun-centered coordinate
system in order to study the sidereal and solar effects separately.  
\end{abstract}
\section{Introduction}
Project GRAND is an extensive air shower array which detects secondary 
muons.  Project GRAND's ability to identify the direction of origin for a 
muon track at a precise time allows for the creation 
of a map in right ascension and declination.  This paper updates previous 
work: \cite{poirier99}, \cite{fields97}, and references therein.    
\section{Experimental Array}
Project GRAND is a proportional wire chamber array located at $86^\circ$W 
and $42^\circ$N.  The array has 64 stations, each  
containing four pairs of orthogonal proportional wire chamber (PWC) planes.  
The four pairs of planes are located above each other with a separation of 
200 mm.  Each plane contains 80 cells with a total active area 
of 1.25 m$^2$ achieving a 
resolution of $0.26^\circ$, on average, in the projected planes. Above the 
bottom pair of planes is a 50 mm thick steel plate.  Electrons interact 
with the plate and shower, scatter, or stop 96\% of the time 
while the muons pass through unaffected 96\% of the time. 
The 64 stations record a total of 
approximately 2000 muons per second.  The data on the muons from the 64 
stations are read in 
parallel into a central data acquisition system at 12 MHz.  
The events are sent to eight separate CPUs in sequence for analysis 
in order to reduce the computer dead time.   
On a new event, the master CPU searches for a slave 
CPU not currently busy analyzing a previous event.
Once information on 900 muons has been collected in a single CPU's 
memory buffer, the data from the buffer are 
written out to magnetic tape in a single record.      
A radio receiver tuned to WWVB in Boulder, 
Colorado provides time information with millisecond precision.  As an 
additional back-up time reference, a one MHz crystal is also used.  Information 
stored on magnetic tape includes the beginning and ending time of each 
record and the angle of the track in both the north/south and east/west 
planes.
\section{Data Analysis}
The data collected are used to generate files containing 
information on the number of muons originating from each 
1$^\circ\times1^\circ$ of right ascension and declination 
during a complete sidereal day onto a 360$\times$110 storage grid.  
From January 1997 through December 2000, 
1052 data files were collected containing information on a total of 
110 billion muons.  In 
order to eliminate possible spurious 
variations in counting rate caused by experimental 
problems (such as a station being offline for repair), 
a smoothness test is imposed on the data files of each day.  

First, the number of muons detected is summed over all declinations for 
each degree of right 
ascension and the average and standard 
deviations of the values calculated.  If the standard 
deviation/average ratio is greater than 2.9\%, then that day's 
data file is not used.  There were 
635 data files containing information on 99 billion muons passed 
the smoothness test. 
%
%
\begin{figure}
\includegraphics[width=8.3cm]{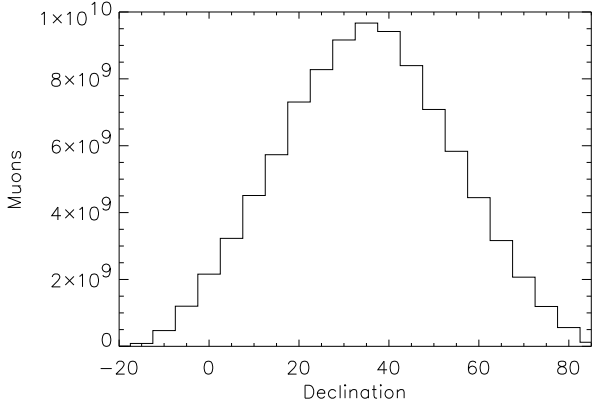}
\caption{A plot of counts vs. declination.  There is a rapid variation 
 caused by angles away from zenith traversing much larger thicknesses of air.
}
\end{figure}
%
\begin{figure}
\includegraphics[width=8.3cm]{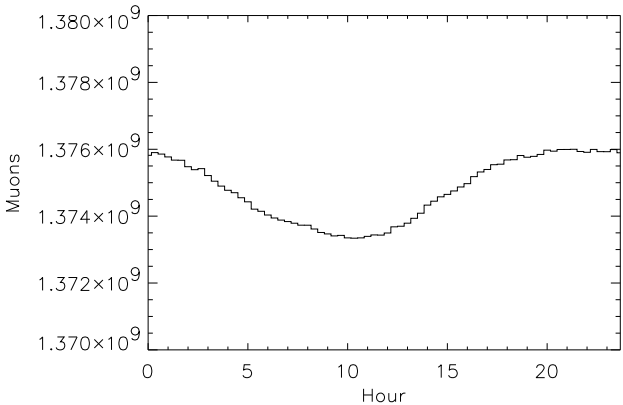}
\caption{A plot of counts vs. right ascension in five degree bins.  The 
dependence is very smooth (notice the surpressed zero).
}
\end{figure}
%
\begin{figure}
\includegraphics[width=8.3cm]{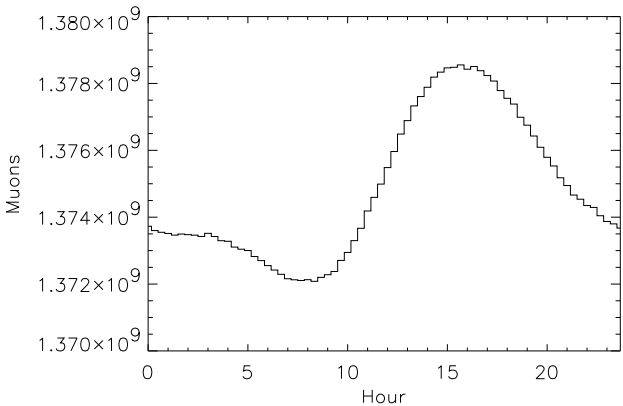}
\caption{A plot of counts versus solar hour-of-day).  The 
dependence is quite smooth (notice the surpressed zero).  
}
\end{figure}
Although Project GRAND has an average muon angular resolution of $0.26^\circ$, the 
secondary muons themselves have a birth-angle relative to the primary and 
are further scattered and deflected as they traverse the atmosphere.  
This degrades 
the resolution for the primary to about $\pm3^\circ$ depending on energy 
\citep{angle}.  
The muon information is summed into $5^\circ$ pixels.  Figure 1 shows 
the number of counts per five degrees of declination.   The counting 
rate has a steep dependence on declination because of the 
much greater acceptance near GRAND's zenith angle ($42^\circ$).
Figure 2 shows the 
number of counts per 5 degrees of right ascension.  The number of counts 
per degree of right ascension is quite smooth; the greatest 
difference shown is only $\pm 8 \times 10^{-4}$ of the average number of counts.   

The pixels were then normalized using equation (1),  
\begin{equation}
Diff=\frac{N(\alpha,\delta)-\langle N(\delta) \rangle}
{\sqrt{\langle N(\delta) \rangle}}
\end{equation} ,
where $N(\alpha,\delta)$ is the number of counts in a given cell and 
$\langle N(\delta) \rangle$ is the average number of counts for 
that declination averaged over all $\alpha$; the denominator is the 
statistical error in this quantity.  
This ratio is the number of standard deviations 
which this cell varies from its average.  

In order to generate a contour map of declination vs. solar time, the 
individual data files are shifted by an amount which  
depends on the number of days past September 23 (the day the sun is at 12 
hr of right ascension).  
Data from September 23 is not 
 shifted while data from later in the year is shifted earlier by 0.9856 times 
the number of days after September 23 (shifted later for earlier days).  
The summed contour map generated using these shifted values also uses the 
normalization of Equation 1; 
Fig. 3 is the projection in half-hour increments of a solar day.  
%
\section{Conclusions}
The $360^\circ$ by $110^\circ$ map of right ascension vs. declination is 
shown in Figure 4 as contours of a given deviation from average 
in units of standard deviations (smoothed).  
The highest deviations are $\geq10 \sigma$.  The 
contour map of declination vs. solar time is shown in Figure 5.  
The greatest 
deviations in this map are $\geq30 \sigma$.  The highest counting rate 
occurs near 16 hours.  Since the map in solar coordinates has the larger 
variations, then perhaps the sidereal variations are caused by a 
solar-linked effect.  
All variations are small compared to the average number of 
62.5 million muons in a $5^\circ$ by $5^\circ$ cell, but 
because of the high statistics obtained, the errors are small even relative 
to the size of, for example, the $\pm 8\times10^{-4}$ variations in 
Figure 2.
\begin{acknowledgements}
Project GRAND is supported through the University of Notre Dame and private 
donations.
\end{acknowledgements}
%
%

 \begin{figure*}[t]
\figbox*{}{}{\includegraphics*[width=17.0cm]{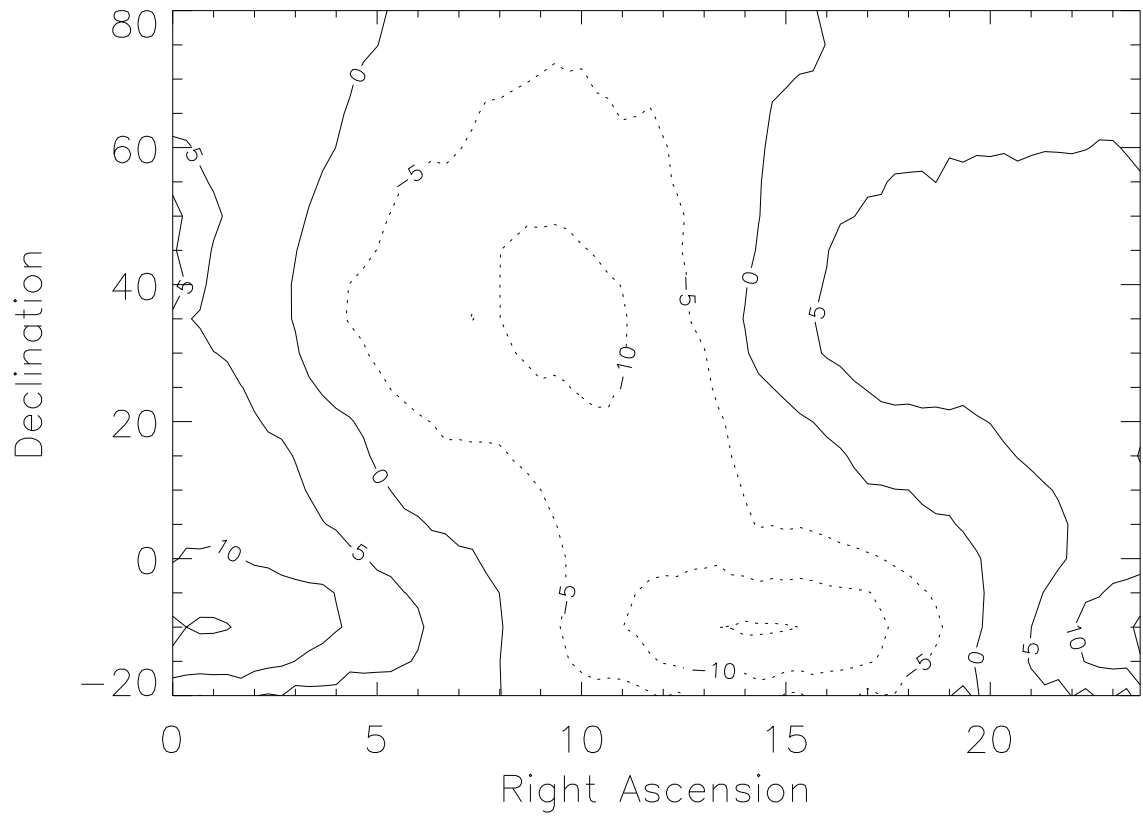}}
\caption
{Deviations from average muon angular distribution.  The contours 
 are the number of standard deviations from average.  The abscissa is in 
 right ascension of a sidereal day.  
 Dashed curves are negative deviations (below 
 average); solid curves are above average muon rate.}
 \end{figure*}
%
 \begin{figure*}[t]
\figbox*{}{}{\includegraphics*[width=17.0cm]{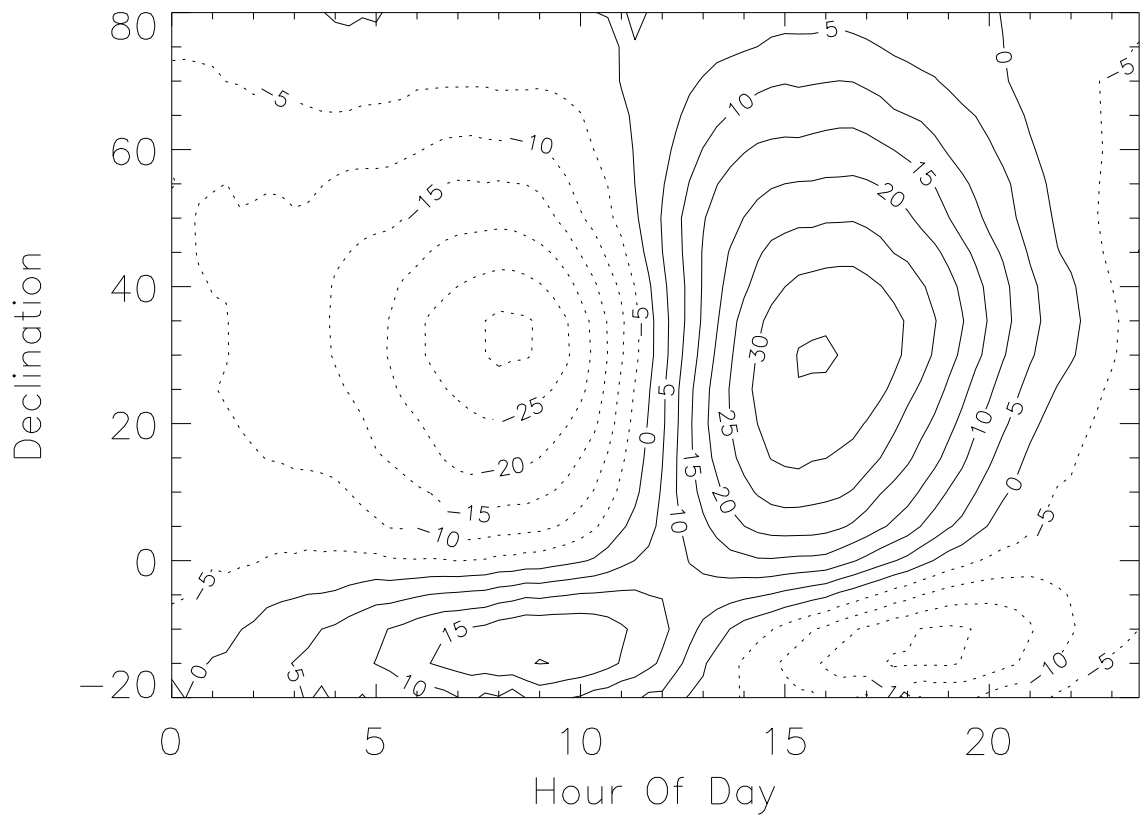}}
 \caption{Deviations from average muon angular distribution.  The contours 
 are the number of standard deviations from average.  The abscissa is in 
 hour of solar day (EST).  Dashed curves are negative deviations (below 
 average); solid curves are above average muon rate.}
 \end{figure*}

\end{document}